# Electron-beam-sustained discharge revisited - light emission from combined electron beam and microwave excited argon at atmospheric pressure

T. Dandl[1], H. Hagn[2], A. Neumeier[2], J. Wieser[3] and A. Ulrich[1]

[1]Physik Department E12, Technische Universität München, James-Franck-Str. 1, 85748 Garching, Germany
[2]Physik Department E15, Technische Universität München, James-Franck-Str. 1, 85748 Garching, Germany
[3]excitech GmbH, Branterei 33, 26419 Schortens, Germany

**Abstract.** A novel kind of electron beam sustained discharge is presented in which a 12keV electron beam is combined with a 2.45GHz microwave power to excite argon gas at atmospheric pressure in a continuous mode of operation. Optical emission spectroscopy is performed over a wide wavelength range from the vacuum ultraviolet (VUV) to the near infrared (NIR). Several effects which modify the emission spectra compared to sole electron beam excitation are observed and interpreted by the changing plasma parameters such as electron density, electron temperature and gas temperature.

## 1. Introduction

The concept of electron beam sustained discharges has been used since the 1970s especially for pumping gas lasers (see e.g. [1, 2]). Laser systems described in these publications typically use an electron beam of ≈ 150 to 250 keV particle energy to ignite gas discharges. They are operated in a pulsed mode with pulse lengths of a few microseconds.

In this publication we present a new kind of electron beam sustained discharge which can be operated in a continuous mode. In a table top setup a low energy electron beam (≈12keV, ≈0,05W) allows us to couple continuous microwave power (2.45 GHz, 30 W maximum) into the target gas. The compact and flexible setup enables measurements with various gases and gas mixtures over a wide pressure range from the millibar region up to 10bar, limited by the stability of the entrance foil for the electron beam. In the experiments described here the pressure was limited to 1.3 bar maximum due to the pressure gauge and the metal bellows pump presently used. Minor modifications of the setup would allow even higher pressures up to 10bar. However, this paper focuses on measurements performed at atmospheric pressure. Operating the electron beam sustained discharge in a continuous mode offers enough time to record high quality optical emission spectra of the plasma with a scanning monochromator over a very broad spectral range from about 60nm [3] up to 1000nm.

In principle, this experiment combines two very different methods of exciting gaseous media: particle beam excitation on the one hand, and discharges on the other. Discharges with their broad field of application in the lighting, coating, and surface treatment industry are commonly used in a low pressure regime, simplifying the ignition and working conditions. Nevertheless, gas discharges at atmospheric pressure gain more and more interest for industry and research, as well. Particularly the ignition process can cause problems in high pressure conditions, since the electrons may not reach enough energy for ionization due to frequent collisions and short mean free path lengths. One way to overcome this problem is to use a particle beam for discharge ignition as it had been done in case of the electron beam sustained discharges described in ref. [1] and [2]. Low energy particle beam excitation (an overview is given in ref. [4] and [5]) can in principle be applied over a very broad pressure range since the particles first gain their energy outside the gas volume and are then guided into a target chamber where they homogeneously excite and ionize the gas volume.

A big difference between the two excitation mechanisms is the shape of the electron energy distribution function (EEDF): in a discharge regime the electrons are cold at the beginning and have to be accelerated until they reach an energy level at which they are able to excite gas atoms or molecules. In this case the first excitation levels of atoms and molecules act like an energy barrier for the electrons, since when reaching this level, the electrons will easily lose their energy by a collision with a gas atom or molecule before gaining again more energy. For particle beam excitation the situation is completely different. In this case, the primary particles are fast when entering the gas volume and they, as well as secondary electrons, are able to excite and ionize target species in multiple collisions before they are cooled down below the energy barrier mentioned above. Compared to discharges, this results in a rather similar respective EEDF at low energies, however showing a long high energy tail (compare ref. [6]), extending out to the particle energy of the incident electron beam.

The effects observed on the emission spectra, which will be described below, are interpreted by the variation of the plasma parameters electron density, electron temperature and gas temperature. A quantitative measurement of these parameters in small plasma volumes (few cubic millimeters in the experiments described here) is difficult, since using e.g. a Langmuir probe might influence the plasma parameters too much. Spectral line broadening of easily accessible optical transitions cannot be used as an alternative since the electron density is too low for using this method. As will be discussed later, the mean electron density for combined electron beam and microwave excitation is roughly comparable to the one for sole electron beam excitation which is estimated to be on the order of $10^{11}$ to $10^{12}$ cm$^{-3}$. Simulations in the case of krypton under similar conditions [7] indicate, that the electron temperature can in principle be varied roughly in the range of 0.5 up to 2 eV for combined excitation.

The setup described here enables us to study and compare three different modes of operation: Sole electron beam excitation, combined electron and microwave excitation, and a self-sustained microwave discharge. The electron beam power can be raised up to 0.1 W in DC operation and in the combined operation additional microwave power up to 0.5W was coupled into the gas target. For a self-sustained discharge this value is about a factor of 10 higher. The self-sustained type of excitation has already been described in ref. [8] and will not be discussed here in detail.

This publication focuses on the combined excitation of argon at atmospheric pressure. After a detailed description of the experimental setup, various effects observed in the optical emission spectra, ranging from the vacuum ultraviolet (VUV) to the near infrared (NIR), are described in detail in chapter 3.

## 2. Experimental setup

### 2.1 Concept

A schematic drawing of the setup is shown in fig. 1. The target cell is made of a standard 40mm diameter (CF40) double cross piece. The electron beam with a particle energy of 12keV is produced in a cathode ray tube (CRT) and enters the gas target through a thin (300nm) ceramic membrane. The gas constantly circulates through a rare gas purifier and the target cell to obtain clean emission spectra. The emitted light is detected by a compact grating spectrometer (Ocean Optics 65000) and a windowless VUV monochromator (Acton Research VM502), respectively. The microwave power is capacitively coupled into the target gas. Important parts of the experimental setup are described in more detail in the following five sections. The basic technique of electron beam excitation of gases used here has already been described in ref. [9] and [5] in the context of vacuum ultraviolet light sources.

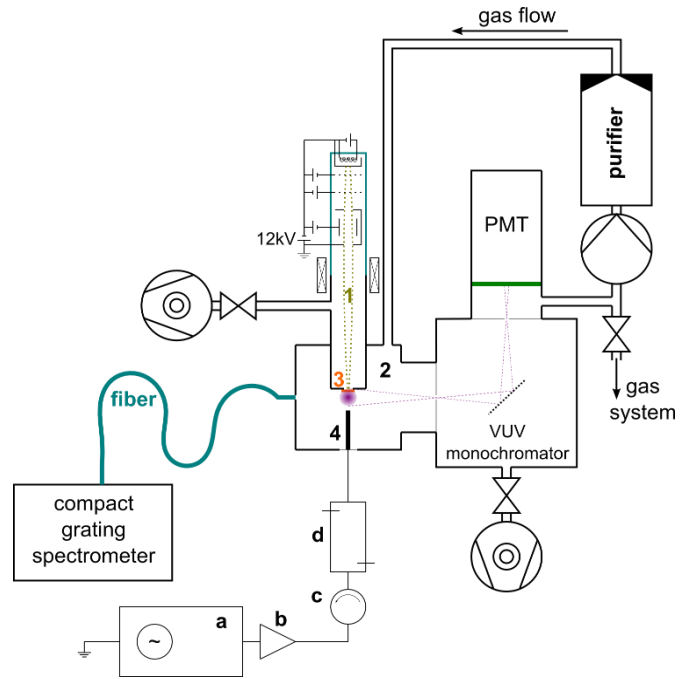

**Fig. 1.** Schematic drawing of the experimental setup. An electron beam (1) is sent into the target chamber (2) through a very thin (300nm) SiN/SiO$_2$ membrane (3). The microwave electrode (4) is mounted on the opposite side. A compact grating spectrometer detects light in the range from about 200nm up to 1000nm. Light of shorter wavelength is detected by a PMT in combination with a VUV monochromator which is included in the gas system to avoid any optical windows. The microwave power is generated (a), amplified (b), sent through a circulator (c) and measured with a directional coupler (d).

### 2.2 Electron beam

The electron beam is formed by a cathode ray tube (CRT, Richardson Electronics) and accelerated to a particle energy of 12keV. A turbo molecular pump keeps the vacuum in the tube in the low $10^{-7}$mbar pressure region. A DC beam current of 10µA maximum is produced and regulated by a computer controlled power supply (Optimare EPU). This results in a power deposition of up to ≈ 0.1 W in the gas target. A higher beam current would destroy the entrance membrane of the beam by overheating. The electron source can also be operated in a pulsed mode with a pulse length down to 100ns. In this case a significantly higher beam current can be used as the membrane can tolerate a higher energy load during the pulses due to its heat capacity, and has the time to cool down between two pulses [10].

As the electron energy of 12keV is quite low, the electrons are stopped in a small volume of the target gas. At atmospheric pressure this volume is of the order of few cubic millimeters [11]. Measurements of the electron beam excited volume are discussed for neon and nitrogen in ref. [12] and are consistent with model calculations with the software Casino [13, 14]. The ionization of the gas caused by the electron beam leads to a mean electron density in the irradiated gas volume of about $10^{11}$ to $10^{12}\,cm^{-3}$ in the case of argon. This value is estimated by calculating the production rate of free electrons by the electron beam including a W-value of 26.4eV (mean energy to form an electron ion pair [15]) and the electron temperature dependent decay rate given by the recombination constant from [16]. A variation of the electron temperature on the order of 0.5 to 2 eV [7] causes only a minor change of the electron density.

Additional microwave power can easily be coupled into the ionized region of the gas target. In the combined operation mode, described in this publication, the electron beam has to be operated during the whole measurement to maintain enough free electrons. Although the electrons are heated significantly by the microwave power in this combined excitation, they do not produce enough free electrons by ionization of argon atoms to overcome the recombination losses. However, using higher microwave power levels, it is possible to ignite a self-sustained microwave discharge in this setup, where the electron beam is just needed for the ignition process and can be switched off afterwards. This has already been described in ref. [8].

The CRT electron gun used here is extended by a stainless steel tube as can be seen in fig. 1. The electron beam is thereby guided to the center of the target cell to allow light detection. Using the focusing electrode of the CRT and two pairs of magnetic coils, the electron beam can be steered exactly through the entrance membrane described in the next paragraph.

### 2.3 Membrane

The evacuated tube of the electron source has to be separated from the target gas. In principle, various types of foils are available for this purpose. Depending on the material and thickness of these foils, a certain minimum electron energy is needed for penetration. For commonly used, pinhole-free metallic membranes, such as titanium foils, this energy is on the order of at least several tens of keV as described in ref. [9] and [5]. For our application the electron energy E should be as low as possible for two reasons: firstly, slow electrons lose their energy in a small gas volume which leads to a high brilliance of light emission and high ionization densities scaling with $1/E^{4.25}$ [5]. And secondly the emission of hard X-rays is avoided. This allows us to build a table top setup for the measurements.

The key technology was the use of very thin (300nm) ceramic membranes, as described in ref. [9]. The membrane used in the setup described here has a size of 0.7 x 0.7 mm² on a 5 x 5 mm² piece of a silicon wafer. Such a membrane can withstand a pressure differential of up to 10bar. The energy loss of 12keV electrons in the membrane is only about 15% [5], which allows a DC beam current of up to ≈ 10µA. A higher beam current can be used in pulsed operation as mentioned above.

It was of great interest for the setup described here whether the membranes can withstand the microwave power and the exposure to a microwave discharge. It turned out that membranes with a thin metallic layer (to enable a better removal of electrons which are backscattered from the gas, charging up the membrane), which are used in several setups, are not suited for this application, since all such membranes broke when igniting the microwave discharge. Membranes without this metallic layer, however, survived hundreds of hours of operation.

### 2.4 Gas system

Since a high gas purity is a very important prerequisite for observing e.g. clean excimer spectra undisturbed by impurities the whole setup described here is based on stainless steel components. Clean gases with a purity of at least "4.8" (99.998% pure) were used and pumped constantly through a rare gas purifier (Mono Torr, Phase II, PS4-MT3) to remove residual impurities from the gas and, above all, impurities which come off the walls of the tubing etc. A metal bellows compressor maintained the gas circulation during measurements. Gas pressure was measured with a capacitive pressure gauge (MKS Instruments, 690A Baratron). Measurements described here were typically performed at atmospheric pressure, but in principle the setup allows us to scan a pressure range from 0 up to 10 bar (limited here to 1.3 bar due to the limit of the pressure gauge presently used and the metal bellows pump). The system was pumped down to $10^{-6}$ mbar prior to gas filling. Note that the vacuum in some parts of the system may not reach this value due to rather thin connection pipes with an inner diameter of 4mm. Therefore the whole system was flushed several times with the new gas.

### 2.5 Microwave generation

Microwave power was generated by an S-Band signal generator (Kuhne Electronic, KU SG 2.45 – 30 A). It delivers a power of up to 30W at 2.45GHz. The generator is internally protected from reflected power by a circulator. A directional coupler allowed us to determine power in forward direction and the power reflected at the target cell separately. From these measurements the power coupled into the system can be derived. The S-band power is capacitively coupled into the target. The microwave electrode forms an antenna inside of the target cell. Note that no efforts were put into optimization of the coupling, as it was not the goal of this study to maximize the power input into the gas, but to observe the effects of the additional power on the optical emission spectra induced by the electron beam. The grounded electrode was formed by the stainless steel extension tube of the electron source. This means that the electron beam was injected into the discharge volume through one of the electrodes. In the combined operation mode, described here, up to ≈ 0.5 W microwave power was coupled into the pre-ionized plasma. Calculating the electron density, corresponding to a plasma frequency of 2.45GHz, gives a value of $7.3\times10^{10}\,cm^{-3}$ which is lower than the average electron density of the gas volume, irradiated by the electron beam. Therefore the microwave is not able to excite the whole pre-ionized volume since it is reflected at the inner part of this plasma in which the electron density exceeds the value of $7.3\times10^{10}$cm$^{-3}$.

## 2.6 Light detection

Light emitted from the target gas was detected by two spectrometers. For the UV-VIS-NIR range (ultraviolet, visible and near infrared) a compact grating spectrometer (Ocean Optics QE65000) was used, which records spectra in the range from 200 to 950 nm in a single exposure. The second one, mainly for the vacuum ultraviolet (VUV) range, consisted of a VUV scanning monochromator (Acton, VM 502) in combination with a photomultiplier tube (PMT). The entrance window of the monochromator was removed (as it is described in ref. [3]). This allowed us to measure light in principle at wavelengths as short as 30nm, thereby being able to observe the continuum radiation from excited helium and neon, and especially here, also the first continuum of argon in its full shape. An entrance window, typically made of MgF$_2$ or LiF, would lead to a short-wavelength cut-off at about 110nm or 105nm, respectively. Since we removed this window, the whole monochromator was filled with the rare gas to be used for the measurement. The rare gases are almost fully transparent for the typical distances in this setup except for regions at and close to the resonance lines. To shift the light to a detectable wavelength region for the PMT, a scintillator was mounted in front of the PMT. The scintillator material was tetraphenyl-butadien. Its optical characteristics are described in ref. [17].

## 3. Experimental Results

### 3.1 Light emission from argon: General aspects

The table top setup described above is well suited for studying three different modes of operation: sole electron beam excitation, a combined excitation of electron beam and microwave power, and a self-sustained microwave discharge. Due to the high pre-ionization of the gas by the electron beam, the ignition of a self-sustained microwave discharge can be achieved with just a few Watts of microwave power at atmospheric pressure [8]. This mode is similar to the classical electron beam sustained discharge described e.g. in ref. [1, 2].

The main focus of this publication lies on a new form of electron beam sustained discharge. The system is operated in a continuously combined electron beam and microwave excitation but without igniting a self-sustained microwave discharge. In this case the electron density is affected mainly by the electron beam current whereas the electron temperature is affected mainly by the microwave power as will be described in the following sections. Therefore the newly developed table top setup allows us to vary these plasma parameters almost independently from each other. Optical emission spectroscopy over a wide wavelength range from the vacuum ultraviolet (VUV) to the near infrared (NIR) is used to interpret the influence of the parameters comparing sole electron beam excitation with the combined excitation. All experiments described here were performed with argon at atmospheric pressure.

Photographs of the light emitting volume are shown in fig. 2 for both excitation methods. The difference in the visible wavelength range can clearly be observed by a color change from a weak violet, in the case of sole electron beam excitation (A), to a bright whitish emission for the combined excitation (B). Furthermore an increase of the light emitting volume (but without really reaching the powered microwave electrode) can be observed. It should be pointed out again, that in case of the combined excitation there is no self-sustained microwave discharge ignited (in contrast to ref. [8] fig. 6C), which is indicated by the absence of any electrode sheath. This means that free electrons, produced by the electron gun in the gas volume are needed to couple microwave power into the gas. Switching off the electron beam leads to a complete loss of light output. This means that the microwave excitation does not produce enough free electrons to compensate the recombination losses.

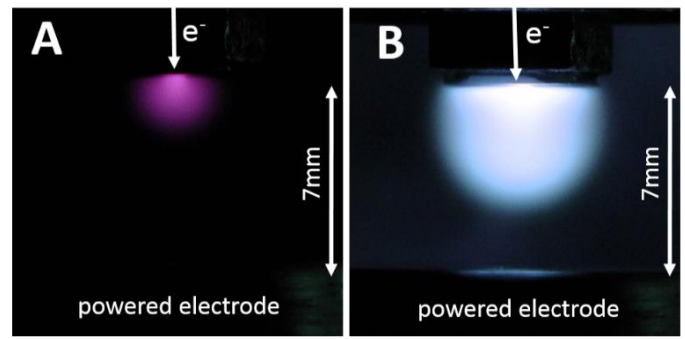

*Fig. 2.* Two photographs showing the light emitting volume of argon at atmospheric pressure for sole electron beam excitation with a beam current of 5µA (A) and combined electron beam (5µA) and 0.5W microwave excitation (B). The tip of the electron gun and the powered electrode for combined microwave excitation are at the same position in both pictures but better visible in B. The absence of an electrode sheath for combined excitation (B; what might look like a sheet is a reflection on the surface of the electrode) shows that there is no self-sustained microwave discharge ignited.

Two overview spectra recorded for 1bar argon are shown in fig. 3. Both spectra are a composition of data recorded with the scanning monochromator and with the compact grating spectrometer respectively. The intensity values were scaled with respect to each other in an overlap region at about 300nm. It is clearly visible that additional microwave power leads to strong modifications in the entire spectral range. In the vacuum ultraviolet (VUV) range a shift in intensity from the so-called second excimer continuum (126nm in argon) to the first continuum located at shorter wavelengths and forming a shoulder on the side of the second continuum as well as to the "classical left turning point" (155nm, LTP: see fig.4) is observed. The shape of the so-called third excimer continuum, located roughly between 170 and 280nm in the case of argon, is also modified. In the ultraviolet, visible and near infrared range (UV-VIS-IR) additional continuum radiation appears. And finally the intensity of the line radiation in the VIS-IR range is modified. These effects will be described in detail below, comparing sole electron beam excitation with combined electron beam and microwave excitation.

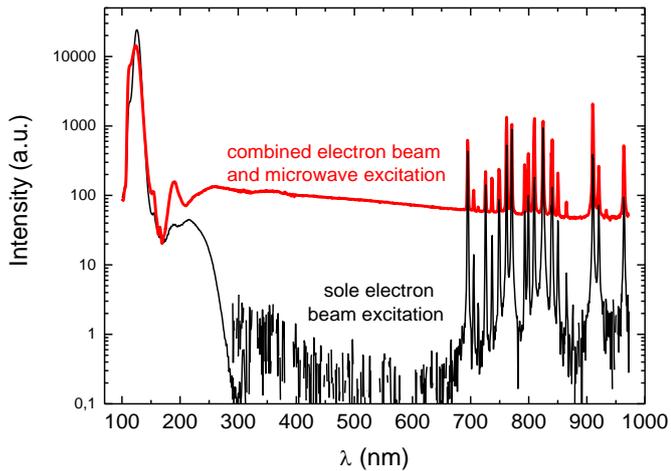

*Fig. 3. Two overview spectra comparing sole electron beam excitation and combined electron beam and microwave excitation of argon at atmospheric pressure.* Both are a composition of spectra recorded with the scanning monochromator and spectra taken with the compact grating spectrometer. Strong modifications of the emitted *light can be observed over the whole range from 100 nm up to 1000 nm. Note, that the spectra are well represented also at the short wavelength end near 100nm because of the windowless spectroscopy described in section 2.6.*

### 3.2 VUV range: First and second excimer continuum

The dominating structure in the VUV range is a very intense broad band radiation formed by the first and second rare gas excimer continua [18, 19]. Excited argon atoms leading to excimer molecule formation are partly produced by direct excitation via collisions with electrons. They can also be formed via recombination and subsequent cascade processes starting with argon ions, which are also produced by collisions with electrons. A schematic overview over the excitation processes and light emitting optical transitions is shown in the potential diagram in fig. 4. Fig.

5A shows the emission spectra from argon at atmospheric pressure for sole electron beam excitation with various beam currents ranging from 0.3 to 7 µA. Very clean excimer continua can be seen without any impurity lines, due to the gas purification system. The intensity of both continua (first and second) is enhanced nearly proportional to the increasing beam current for low values as can be seen in fig 5B. For higher beam currents the enhancement weakens a bit due to a not perfect beam focus. This means for low beam currents the whole electron beam can be steered through the small entrance membrane. As the beam diameter increases a bit with increasing current the outer parts of the beam are cut away by the edges of the silicon wafer. Earlier measurements with a narrower beam profile had shown a perfectly linear relation over the whole range.

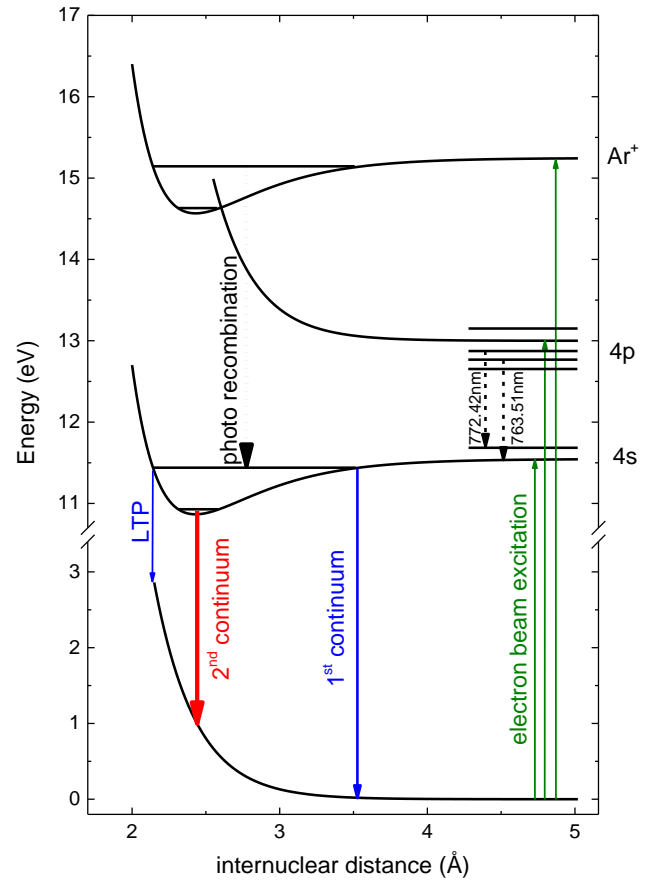

*Fig. 4. Schematic potential diagram of argon showing the main excitation processes by electron collisions. The optical transitions leading to the first and second excimer continuum in the VUV range are also indicated as well as the two atomic lines which are discussed in the text.*

The rising intensity of both continua means that a higher electron density in the gas caused by a higher beam current leads to a proportionally higher formation of excimer molecules. However, the intensity ratio between the two continua changes also a bit (≈5%) which can be seen in fig. 5B. The slightly stronger increase of the first excimer continuum intensity with beam current indicates a correspondent increase of the gas temperature [20].

Fig 6A shows the same spectral range but now the electron beam current is kept constant at a low value of 0.3 µA to keep the electron density low and to enable the microwave power to act on a large part of the electron beam excited volume (compare section 2.5). Microwave power coupled into the gas is increased stepwise from 0.04 up to 0.44W. Here a decrease in intensity of the second continuum and an increase of the first continuum can be observed. This indicates a redistribution of the vibrational excited excimer levels schematically shown in the potential diagram in fig.4. The optical transitions leading to the first and second continuum are also indicated in this figure. An intensity shift from the second to the first continuum, which can be observed in the spectra, shows clearly a redistribution of the population from energetically low lying vibrational states to energetically higher lying states. This population redistribution shows that newly formed excimer molecules are prevented from a relaxation to energetically lower lying states, which happens mainly due to collisions with electrons. Therefore this effect shows the increase of the electron temperature (and gas temperature) with increasing microwave power.

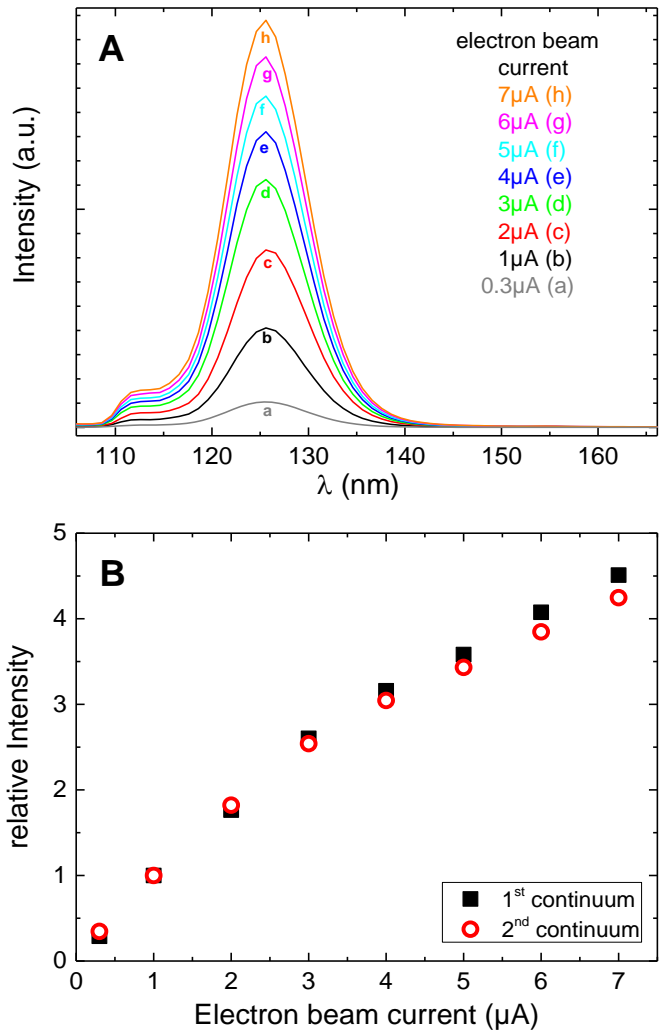

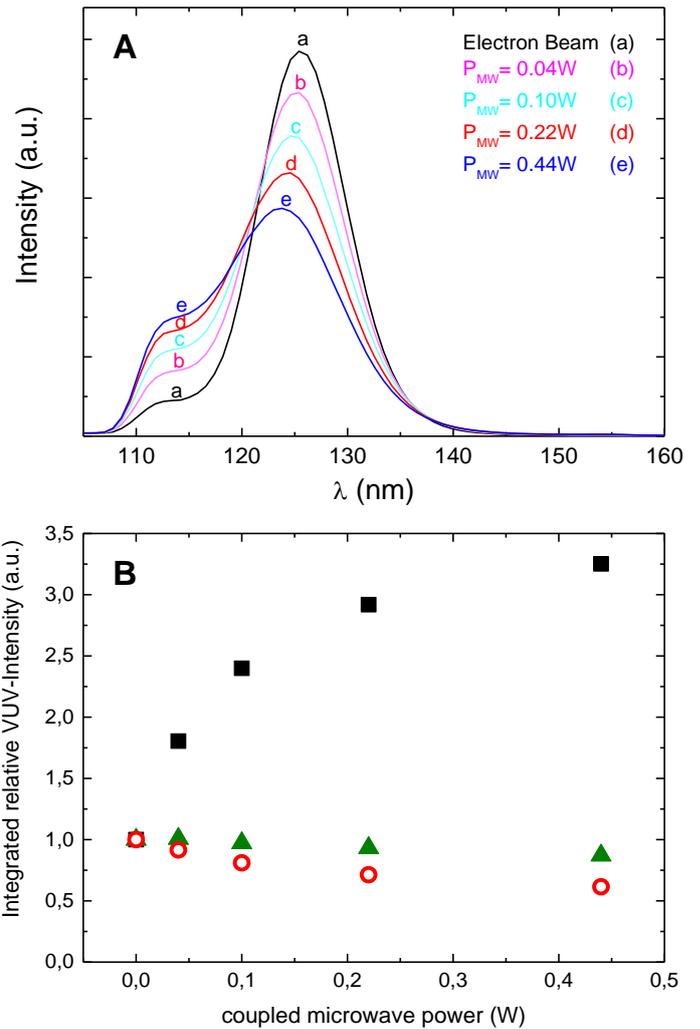

*Fig. 5.* First and second excimer continuum of argon at atmospheric pressure for electron beam excitation are shown for various electron beam currents (A). Integrated values relative to the 1µA measurement are given in (B) for the first continuum ($\lambda<115$nm, black squares) and the second continuum ($\lambda>115$nm, red circles). The non-linearity is mainly due to losses because of a not perfect beam focus. The slightly stronger rise of the first continuum is a hint to a slightly rising gas temperature.

*Fig. 6.* First and second excimer continuum of argon at atmospheric pressure are shown for combined electron beam and microwave excitation (A). Integrated values relative to the 0.3µA measurement without microwave excitation are given in panel (B) for the first continuum ($\lambda<115$nm, black squares), the second continuum ($\lambda>115$nm, red circles) and the overall VUV-range (green triangles). The decreasing trend of the second continuum and the rising of the first continuum are clearly visible. The overall VUV intensity shows a slightly decreasing trend.

The total electron density, however, should stay rather constant for rising microwave power, since the overall intensity integrated over both continua is not enhanced. This indicates that for the combined excitation mode the microwave does not couple enough power into the gas to excite or even ionize a large number of argon atoms from the ground state which would lead to a higher production rate of excimer molecules. In contrary, even a slight decrease in intensity can be observed as shown in fig. 6B. This trend may have several reasons. Higher absorption losses of the short wavelength light on its long way through the gas filled monochromator might be one of them, since there is an enhanced light production close to the resonance lines of atomic argon. But also partial destruction of excimers due to a higher temperature or a reduced production of excimers due to re-excitation of argon atoms, as well as particularly a decreased recombination rate are possible explanations for the slight VUV-intensity decrease.

## 3.3 Third excimer continuum in the deep UV range

Fig. 7 shows the third excimer continuum which is located roughly between 170 and 250nm in the case of argon. There are still investigations to complete the understanding of some features of this continuum [21-25]. According to ref. [23] a combination of several optical transitions from singly and doubly ionized molecules and clusters could form this continuum radiation. The dominant parts peak at 188 and 199nm for doubly ionized molecules and at 176, 212, 225 and 245nm for singly ionized molecules. The spectrum shown in black in fig. 7 was recorded with sole electron beam excitation of argon at atmospheric pressure and may be a combination of all those parts. Adding and rising the microwave power from 0.04 to 0.22W, only leads to an amplification of the parts assigned to doubly ionized molecules (188 and 199nm), whereas the other structures are decreased in intensity. Note, that the rising intensity of the 0.22W spectrum at longer wavelengths is due to an overlap with the next continuum radiation feature, which is discussed in the following section. The total intensity of the third continuum remains constant for additional 0.04W and 0.1W microwave power and only a shift in intensity from transitions belonging to singly ionized molecules to the ones belonging to doubly ionized molecules can be observed. This might indicate a reduced recombination rate from doubly to singly ionized molecules due to a higher electron temperature.

In more detail, the emission of the different parts of the third continuum happens as a cascade of different transitions as it is discussed in detail in [23] (see flow diagram in fig. 10 in this reference). If the doubly ionized molecules (emitting at 188nm and 199nm) undergo a recombination process leading to a singly ionized molecule, they can contribute to the emission features at 176, 212, 225 and 245nm. Since an intensity shift from the last four transitions to the first two occurs, the recombination process in this cascade seems to be suppressed. The exact cross section for this process is not known, but it should show an inverse dependence on the electron and gas temperature, which is typical for a recombination process. Therefore, the behavior of the third continuum with rising microwave power also indicates a rising electron and gas temperature. The fact that the integrated intensity of the third continuum remains constant (trace a, b and c in fig. 7), shows again that the microwave power is too low to populate high lying excited or even ionic states from the ground state.

## 3.4 Continuum radiation in the UV-VIS range

The combined excitation mode leads to an additional spectral feature which is shown in fig. 8 and could not be observed for sole electron beam excitation, so far. To obtain the spectra shown here, the electron beam was reduced to the lowest reachable limit (< 0.1 µA) and was therefore no longer quantitatively measurable with the power supply used here. This was done to observe the strongest effect possible caused by the additional microwave power without igniting a self-sustained discharge. The new broadband structure starts in the deep UV range partly overlapping with the third continuum.

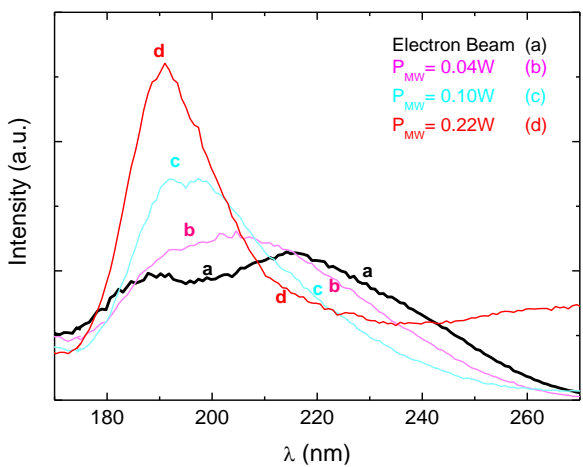

***Fig. 7.*** *Third excimer continuum emission of argon at atmospheric pressure. Sole electron beam excitation (black curve, a) is compared with a combination of electron beam excitation and microwave excitation at various power levels (b: 0.04 W – c: 0.10 W – d: 0.22 W). A shift in intensity towards the short wavelength parts of this continuum can clearly be seen with increasing microwave power. For higher power levels (0.22 W) the long wavelength part overlaps with an additional continuum discussed in section 3.4.*

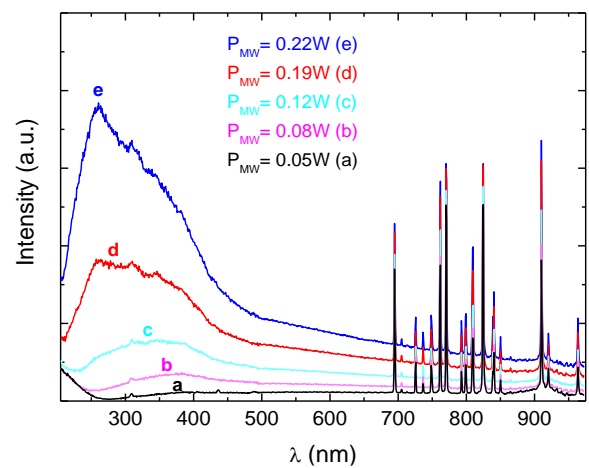

***Fig. 8.*** *UV-VIS-continuum of argon at atmospheric pressure for combined electron beam and microwave excitation. The electron beam current was kept below a measurable value of ≈ 0.1µA. For rising microwave power a strong increase in intensity can be observed especially at the short wavelength side of this continuum radiation. Near the short wavelength cut-off of this emission feature it partly overlaps with the third excimer continuum discussed in section 3.3.*

Such a continuous light emission can originate from Bremsstrahlung of electrons on atoms, molecules and ions (free-free) or from photo-recombination of electrons with ions (free-bound). It has been widely studied and is still being studied in recent measurements (compare e.g. ref. [26-32]). In principle the continuum can be a combination of both processes with varying contributions depending on the plasma parameters. In our case the origin of this continuum cannot fully be assigned to one or the other mechanism, since so far we were not able to measure electron temperature and electron density quantitatively. Nevertheless, the rather low electron temperature in our experiments (compared to nearly 5eV photon energy at the peak intensity at ≈ 260nm) indicates the continuum, especially in the wavelength range shown in fig.8, most likely to be dominated by photorecombination. Ionic argon molecules could in this case photo-recombine with free electrons into vibrationally excited excimer molecules. This provides an energy of almost 4eV (fig. 4). The one missing electronvolt for the 5eV corresponding to the peak intensity of the continuum has to be provided by the thermal energy of the electron which roughly confirms the simulation of electron temperature in ref. [7]. Further experiments, especially with pulsed excitation, shall help for a better understanding.

We have not observed this continuum for sole electron beam excitation so far, not even shifted to longer wavelengths (optical emission spectra up to a wavelength of 3.5μm for electron beam excited argon will be published in a forthcoming paper), which is a further argument against a Bremsstrahlung continuum. Nevertheless, a photo recombination continuum could also be expected in the case of sole electron beam excitation. We cannot proof the following reason for the absence of this continuum for electron beam excitation, but we want to suggest it as a possible explanation: the starting point for photorecombination are ionized argon atoms and molecules. The related potential curve is indicated in fig.4. Similar to the case of neutral excimers (see section 3.2) most of the molecular ions will exist in vibrationally relaxed states for sole electron beam excitation (and therefore very low electron temperatures). In this case the dissociative recombination (indicated by the dissociative potential curve connecting the $Ar^+$-curve with the 4p-states in fig. 4.) might be the dominant recombination process. It has to be admitted, that this is difficult to quantify since no photo recombination cross sections for ionized argon molecules are available in literature. As already described in section 3.2, heating up the gas by additional microwave power leads to a redistribution of population of molecular states to energetically higher lying states. This might shift the balance of the recombination process from a dissociative recombination to a photo recombination, which can be observed as the continuum emission shown in fig. 8 for the addition of microwave power, only.

Independent from the origin of the continuum, it is correlated with the rising electron temperature with rising microwave power. The peak intensity of the continuum is shifted to shorter wavelengths and therefore higher photon energies with increasing microwave power (see fig. 8). Since the temperature of atoms or molecules will not contribute a significant amount of energy to the ongoing process (neither Bremsstrahlung nor photo recombination), this spectral effect also shows the rising electron temperature very nicely.

### 3.5 Atomic Line Radiation in the VIS-NIR Range

Line intensity ratios are often used to determine plasma parameters, for example via lines of a small amount of hydrogen in the gas. There also exist measurements with atomic argon lines as described e.g. in ref. [34]. If one plasma parameter, electron density or electron temperature, is known, the variation of the intensity ratio of line radiation can be used to derive the other parameter. Since, so far, we have not the possibility to measure the plasma parameters directly to cross check these values, we will use in the following the behavior of argon line radiation to interpret the relative variation of the plasma parameters.

As an example, fig. 9 shows two atomic lines emitted from 1bar argon for the two different excitation methods discussed above and recorded with the low resolution spectrometer. The underlying continuum radiation described in the previous section was subtracted from each spectrum to show only the variation of the line intensity. The line radiation is directly connected to the excimer continuum radiation as it is one way to populate the argon 4s states, (described above and indicated in fig 4). In the case of sole electron beam excitation this connection can clearly be seen in the spectra in fig. 9a, which shows as an example an expanded view of two lines (763,51 and 772,42nm) originating from the two different excited 4p states $4p[3/2]_2$ (13.17eV) and $4p[1/2]_1$ (13.33eV) [35] in Racah notation. The behavior of the line intensity depending on the electron beam current is nearly identical to the behavior of the VUV continua discussed above and shown in fig. 5B. The uniform increase of all line intensities with rising electron beam current, independent from the initial state of the optical transition, shows that the increasing electron density happens without a redistribution of excited states involved and therefore also without a significant variation of electron temperature.

Fig. 9B shows the same spectral range for a fixed beam current of 0.3μA and a stepwise increased additional microwave power. In contrast to the sole electron beam excitation the additional microwave power causes a very different intensity increase of lines originating from different 4p states. As it was discussed in section 3.2, additional microwave power is not able to excite argon atoms from the ground state. Therefore the increase of atomic argon line intensity with rising microwave power happens most likely via re-excitation of argon atoms from the 4s states to the 4p states (compare potential scheme in fig. 4). In this case an increasing electron temperature should lead to a higher population of energetically lower lying 4p states (since the needed excitation energy is lower) and therefore also to a stronger amplification of optical transitions originating from the lower lying states. Although this trend can be observed for the two lines shown in fig. 9B, the extremely diverse intensity increase of the lines can hardly be explained only by the different excitation energy (ΔE≈0.16eV) of the upper states. Other lines (not shown in fig.9) which have an excitation energy between that of the two lines discussed above show an even higher intensity increase than the 763.51nm line. Therefore this excitation energy effect should exist, but it is overlaid with collisional population transfer within the 4p states. Rate constants for gas temperature dependent collisional population transfer within the 4p states are tabulated in ref. [36].

In this publication we do not want to provide a complete model for the population of the different 4p states and the population transfers between these states. But even without a detailed model, the analysis of the two argon lines in fig. 9 can be used to show the influence of electron beam and microwave power on the plasma parameters.

## 4. Conclusion

A table top setup for studying electron beam sustained discharges, operated for the first time in a continuous mode, has been described. A 12keV electron beam was combined with a 2.45GHz microwave excitation. First observations in the spectral range from 100 to 1000nm have been described and discussed. Significant spectral modifications were observed due to varied electron density, electron temperature and gas temperature. These modifications of the emission spectra have shown that the electron beam current affects mainly the electron density, whereas the microwave power affects mainly electron- and gas temperature of the excited gas volume. As could be expected, the inverse, heating by the electron beam, as well as a rising electron density by the microwave power (due to e.g. a reduced recombination with increasing electron temperature) just play a minor part. Therefore the two different excitation methods make it possible to perform measurements with different combinations of the plasma parameters since the electron beam current and the microwave power can be adjusted independently from each other. Further experiments including pulsed excitation will provide a better understanding of the observed effects and will allow to measure the plasma parameters quantitatively (similar to measurements performed in xenon [37]). Even higher excitation frequencies would also be interesting to be used, since this would allow acting on areas in the electron beam excited volume with higher electron densities.

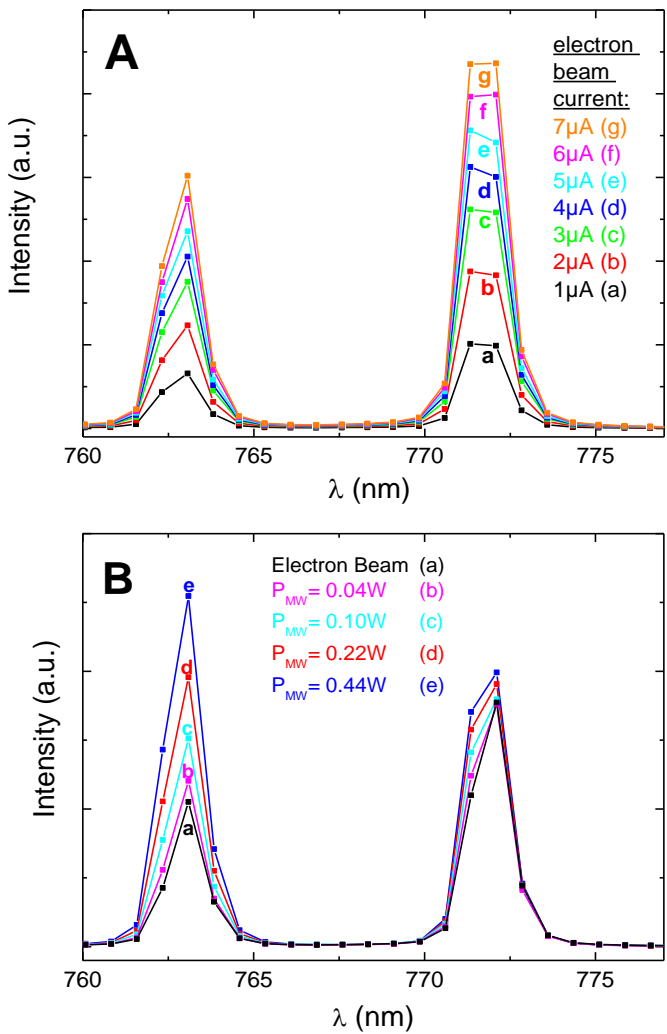

*Fig. 9.* *Two series of measurements are shown comparing the effect of the electron beam current and the microwave power on the atomic line radiation in the visible spectral range. An expanded view for two neighboring lines is shown as an example. Spectra for sole electron beam excitation with varied beam current are shown in (A). The line intensity depending on the electron beam current behaves similar to the excimer continua intensity shown in fig. 5B. Spectra for a fixed electron beam current of 0.3 µA and a variable additional microwave power are shown in (B). The different amplification of the two lines due to increasing gas temperature is clearly visible.*

### Acknoledgement

The authors acknowledge financial support by the Bundesministerium für Bildung und Forschung (BMBF) contract number 13N9528 (SAFE-INSIDE) and 13N11376 (DIVE) and the Maier-Leibnitz-Laboratorium München (MLL).

### References

[1] L. A. Newman, T. A. Detemple, Appl. Phys. Lett. **27**, 678 (1975)
[2] N. W. Harris, F. Oneill, W. T. Whitney, Rev. Sci. Instrum. **48**, 1042 (1977)
[3] A. Fedenev *et al.*, J. Phys. D **37**, 1586 (2004)
[4] A. Ulrich, Laser Part. Beams **30**, 199 (2012)
[5] A. Ulrich *et al.*, Eur. Phys. J. Appl. Phys. **47**, 22815 (2009)
[6] T. J. Moratz, T. D. Saunders, M. J. Kushner, J. Appl. Phys. **64**, 3799 (1988)

[7] G. Zvereva, Opt. Spectrosc. **108**, 4 (2010)
[8] T. Dandl *et al.*, Europhys. Lett. **94**, 53001 (2011)
[9] J. Wieser *et al.*, Rev. Sci. Instrum. **68** 1360 (1997)
[10] A. Morozov *et al.*, Eur. Phys. J. D **48**, 383 (2008)
[11] S. Valkealahti, J. Schou, R. M. Nieminen, J. Appl. Phys. **65**, 2258 (1989)
[12] A. Morozov *et al.*, J. Appl. Phys. **100**, 093305 (2006)
[13] CASINO, http://www.gel.usherbrooke.ca/casino/
[14] P. Hovington, D. Drouin, R. Gauvin, Scanning **19**, 1 (1997)
[15] D. Combecher, Radiat. Res. **84**, 189 (1980)
[16] Y. J. Shiu, M. A. Biondi, Phys. Rev. A **17**, 868 (1978)
[17] W. M. Burton, B. A. Powell, Appl. Opt. **12**, 87 (1973)
[18] Y. Tanaka, J. Opt. Soc. Am. **45**, 710 (1955)
[19] Y. Tanaka, A. S. Jursa, F. J. Leblanc, J. Opt. Soc. Am. **48**, 304 (1958)
[20] R. Prem, Diplomarbeit, TU München Physik Department e12 (1994)
[21] A. B. Treshchalov, A. A. Lissovski, Eur. Phys. J. D **66**, 1 (2012)
[22] A. A. Lissovski, A. B. Treshchalov, Phys. Plasmas **16**, 123501 (2009)
[23] J. Wieser *et al.*, Opt. Commun. **173**, 233 (2000)
[24] A. M. Boĭchenko *et al.*, Quantum Electron. **23**, 3 (1993)
[25] H. Langhoff, Opt. Commun. **68**, 31 (1988)
[26] R. R. Johnston, J. Quant. Spectrosc. Radiat. **7**, 815 (1967)
[27] U. Bauder, J. Appl. Phys. **39**, 148 (1968)
[28] A. T. M. Wilbers *et al.*, J. Quant. Spectrosc. Radiat. **45**, 1 (1991)
[29] D. Schlüter, Zeitschrift für Physik **210**, 80 (1968)
[30] L. G. D'Yachkov, Y. K. Kurilenkov, Y. Vitel, J. Quant. Spectrosc. Radaiat. **59**, 53 (1998)
[31] J. Park *et al.*, Phys. Plasmas **7**, 3141 (2000)
[32] A. B. Treshchalov, A. A. Lissovski, J. Phys. D Appl Phys **42**, 245203 (2009)
[33] X. M. Zhu *et al.*, J. Phys. D **42**, 142003 (2009)
[34] A.R.Striganov, N.S. Sventitskii, *Tables of spectral lines of neutral and ionized atoms* (New York, IFI/Plenum, 1968)
[35] T. D. Nguyen, N. Sadeghi, Phys. Rev. A **18**, 1388 (1978)
[36] G. Ribitzki *et al.*, Phys. Rev. E **50**, 3973 (1994)